\documentclass[aps,preprintnumbers,superscriptaddress]{revtex4}

\usepackage{graphicx}
\usepackage{bm}
\def\gappeq{\mathrel{\rlap {\raise.5ex\hbox{$>$}}
{\lower.5ex\hbox{$\sim$}}}}

\def\lappeq{\mathrel{\rlap{\raise.5ex\hbox{$<$}}
{\lower.5ex\hbox{$\sim$}}}}

\def\ga{\mathrel{\raise.3ex\hbox{$>$\kern-.75em\lower1ex\hbox{$\sim$}}}}
\def\la{\mathrel{\raise.3ex\hbox{$<$\kern-.75em\lower1ex\hbox{$\sim$}}}}
\def\gev{{\rm \, Ge\kern-0.125em V}}
\def\tev{{\rm \, Te\kern-0.125em V}}
\def\beq{\begin{equation}}
\def\eeq{\end{equation}}

\def\m12{m_{1\!/2}}

\def\tb{\tan\beta}

\def\APJ#1#2#3{Ap. J. {\bf #1}, #2 (#3)}

\newcommand{\mgaugino}{M_{1/2}}

\newcommand{\Omegachi}{\Omega_{\chi}}
\newcommand{\eg}{{\em e.g.}}

\newcommand{\km}{{\rm km}}
\newcommand{\cm}{{\rm cm}}
\newcommand{\yr}{{\rm yr}}
\newcommand{\s}{{\rm s}}

\begin{document}

\preprint{MIT--CTP--3199, UCI--TR--2001--32, CERN--TH/2001-297,
          hep-ph/0111295, Snowmass P3-09}

\title{Particle and Astroparticle Searches for Supersymmetry}



\author{Jonathan L.~Feng}
\email[]{jlf@mit.edu}
\affiliation{Center for Theoretical Physics,
             Massachusetts Institute of Technology,
             Cambridge, MA 02139, USA}
\affiliation{Department of Physics and Astronomy, 
             University of California, Irvine, CA 92697, USA}
\author{Konstantin T.~Matchev}
\email[]{Konstantin.Matchev@cern.ch}
\affiliation{Theory Division, CERN,
             CH--1211, Geneva 23, Switzerland}
\author{Frank Wilczek}
\email[]{wilczek@mit.edu}
\affiliation{Center for Theoretical Physics,
             Massachusetts Institute of Technology,
             Cambridge, MA 02139, USA}

\date{November 22, 2001}

\begin{abstract}
Supersymmetry may be discovered at high energy colliders, through low
energy precision measurements, and by dark matter searches.  We
present a comprehensive analysis of all available probes in minimal
supergravity. This work extends previous analyses by including the
focus point branch of parameter space and the full array of promising
indirect dark matter searches. We find that particle and astrophysical
searches underway are highly complementary: each separately provides
only partial coverage of the available parameter space, but together
they probe almost all models.  Cosmology does {\em not} provide upper
bounds on superpartner masses useful for future colliders.  At the
same time, in the cosmologically preferred region, if supersymmetry is
to be observable at a 500 GeV linear collider, some signature of
supersymmetry must appear {\em before} the LHC.
\end{abstract}

\maketitle


Weak-scale supersymmetry has consequences for a wide variety of
particle physics experiments.  At the same time, it is now
well-established that luminous matter makes up only a small fraction
of the mass of the observed universe~\cite{Primack:2000dd}.
Neutralinos in supersymmetry are well-motivated candidates to provide
much or all of the non-baryonic dark matter. While it is conceivable
that all standard model superpartners are unstable, decaying, for
example, into gravitinos or axinos, in many model frameworks, a
neutralino is the lightest supersymmetric particle (LSP) and stable.
Further, neutralino thermal relic densities of the desired amount are
remarkably generic in supergravity models~\cite{neutralinos}.  If
neutralinos make up a significant portion of the halo dark matter,
astrophysical searches for dark matter open up many additional avenues
for the discovery of supersymmetry.

Here we summarize results of two studies of current and near future
prospects for supersymmetry searches~\cite{Feng:2000gh,Feng:2001zu}.
We consider minimal supergravity, a simple framework that is
conservative in the sense that many low energy signals generic in
supersymmetry are suppressed.  This work extends previous analyses in
two ways.  First, we include the focus point branch of supersymmetry
parameter space~\cite{FP}, previously neglected.  As we will see,
neutralino dark matter in the focus point region has novel properties,
with strong, positive implications for dark matter searches.  Second,
we consider not only collider experiments, low energy probes, and
direct dark matter searches, but also the full array of indirect dark
matter searches.  We find that the various searches are highly
complementary, and this comprehensive approach leads to strong
conclusions that may be reached only by considering all available
experiments.

The signals we consider, the projected sensitivities, and the
experiments likely to achieve them, are listed in
Table~\ref{table:comp}.  Conventional particle physics probes include
searches at LEP and the Tevatron, the improved measurement of the the
$B\rightarrow X_s\gamma$ branching ratio at $B$ factories, as well as
the projected final sensitivity of the Brookhaven $g_\mu-2$
experiment.  On the astrophysics side, we consider the projected reach
of detectors searching for direct dark matter interactions, including
the {\tt CDMS}~\cite{cdms}, {\tt CRESST}~\cite{cresst} and {\tt
GENIUS}~\cite{genius} experiments.

We also include the most promising indirect dark matter searches.  In
the next five years, an astounding array of experiments will be
sensitive to potential neutralino annihilation products. These include
under-ice and underwater neutrino telescopes ({\tt
AMANDA}~\cite{amanda}, {\tt NESTOR}~\cite{nestor}, {\tt
ANTARES}~\cite{antares}), which will be sensitive to muons produced by
neutrinos from relic annihilations in the cores of the Sun and
Earth~\cite{neutrinos}; atmospheric Cherenkov telescopes ({\tt
STACEE}~\cite{stacee}, {\tt CELESTE}~\cite{celeste}, {\tt
ARGO-YBJ}~\cite{argo}, {\tt MAGIC}~\cite{magic}, {\tt
HESS}~\cite{hess}, {\tt CANGAROO}~\cite{cangaroo}, {\tt
VERITAS}~\cite{veritas}), and space-based $\gamma$ ray detectors ({\tt
AGILE}~\cite{agile}, {\tt AMS}/$\gamma$~\cite{amsgamma}, {\tt
GLAST}~\cite{glast}), which may detect gamma rays from the galactic
center~\cite{photons}; and anti-matter/anti-particle experiments ({\tt
PAMELA}~\cite{pamela}, {\tt AMS}~\cite{ams}), which may detect
positrons~\cite{positrons}, antiprotons~\cite{antiprotons} and
antideuterium~\cite{antideuterium} from the galactic halo.  In many
cases, these experiments will improve current sensitivities by several
orders of magnitude.  The search for supersymmetric dark matter is
reviewed in a number of excellent articles~\cite{indirectreviews}; see
also Refs.~\cite{RDrecent,direct,recent} and the bibliographies of
Refs.~\cite{Feng:2000gh,Feng:2001zu}.

\begin{figure}[t]
\begin{minipage}[t]{0.49\textwidth}
\includegraphics[height=2.3in]{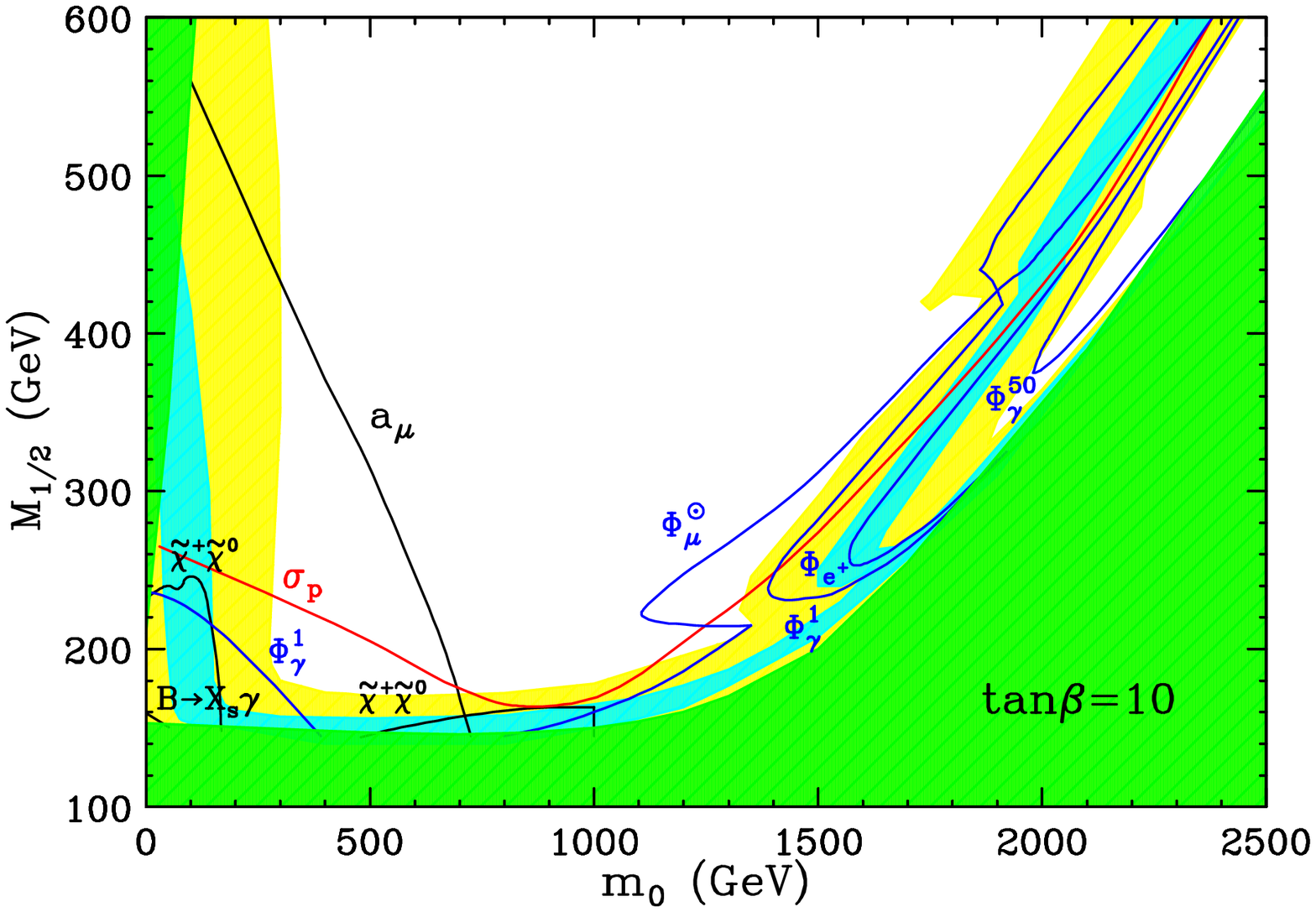}
\end{minipage}
\hfill
\begin{minipage}[t]{0.49\textwidth}
\includegraphics[height=2.3in]{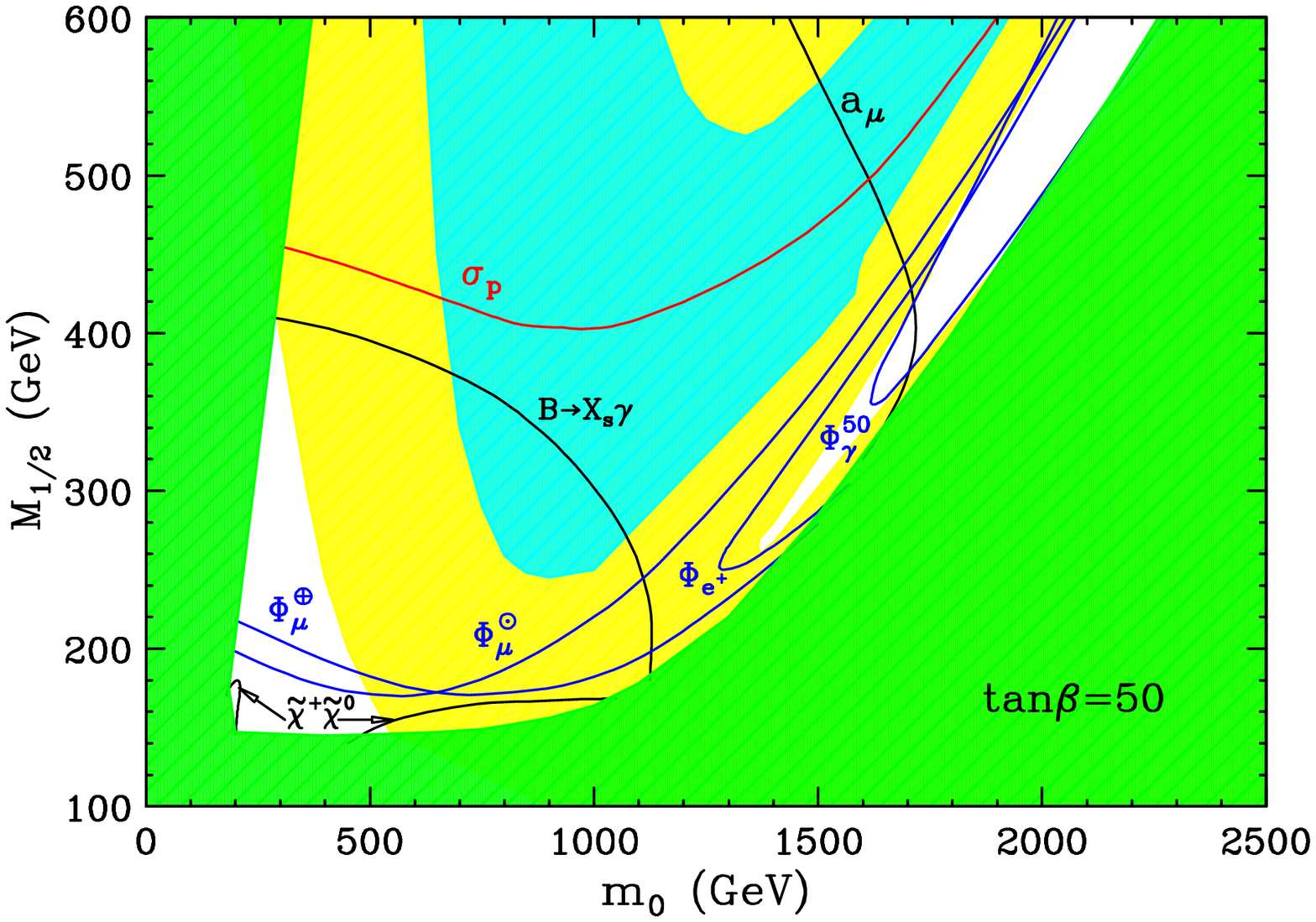}
\end{minipage}
\caption[]{Estimated reaches of various high-energy collider and
low-energy precision searches, direct dark matter searches, and
indirect dark matter searches before the LHC begins operation, for
$\tb=10$ (left) and $\tb=50$ (right).  The projected sensitivities
used are given in Table~\ref{table:comp}. The darker shaded (green)
regions are excluded by the requirement that the LSP be neutral (left)
and by the LEP chargino mass limit (bottom and right).  We have also
delineated the regions with potentially interesting values of the LSP
relic abundance: $0.025\le \Omegachi h^2 \le 1$ (light-shaded, yellow)
and $0.1 \le \Omegachi h^2 \le 0.3$ (medium-shaded, light blue).  The
regions probed extend the curves toward the forbidden regions.  The
dark matter reaches are not modulated by the thermal relic density.
Bounds from photons from the galactic center are highly halo
model-dependent; we assume a moderate halo profile parameter $\bar{J}
= 500$ \cite{Bergstrom:1998fj}. (From Ref.~\cite{Feng:2001zu}.)}
\label{fig:reach}
\end{figure}

\begin{table}[ht!]
\caption[]{Constraints used in Fig.~\ref{fig:reach}, and experiments
likely to reach these sensitivities by 2006.
\label{table:comp}
}
\begin{tabular}{||l|l|l|l||}
\hline
 Observable 
  & Type
   & Bound
    & Experiment(s)   \\  \hline
 $\tilde{\chi}^+ \tilde{\chi}^-$
  & Collider
   & $m_{\tilde{\chi}}^{\pm} > 100~\gev$
    & LEP: {\tt ALEPH, DELPHI, L3, OPAL}  \\ 
 $\tilde{\chi}^{\pm} \tilde{\chi}^0$
  & Collider
   & from~\cite{trileptons}
    & Tevatron: {\tt CDF, D\O\ }  \\ 
 $B \to X_s \gamma$
  & Low energy
   & $|\Delta B(B\rightarrow X_s\gamma)| < 1.2\times 10^{-4}$
    & {\tt BaBar, BELLE}     \\
 Muon MDM
  & Low energy
   & $|a_{\mu}^{\text{SUSY}}| < 8 \times 10^{-10}$
    & Brookhaven {\tt E821}  \\
 $\sigma_{\text{proton}}$
  & Direct DM
   & from~\cite{Schnee:1998gf,Bravin:1999fc}
    & {\tt CDMS, CRESST, GENIUS} \\
 $\nu$ from Earth
  & Indirect DM
   & $\Phi_{\mu}^{\oplus} < 100~\km^{-2}~\yr^{-1}$
    & {\tt AMANDA,NESTOR,ANTARES} \\
 $\nu$ from Sun
  & Indirect DM
   & $\Phi_{\mu}^{\odot} < 100~\km^{-2}~\yr^{-1}$
    & {\tt AMANDA,NESTOR,ANTARES} \\
 $\gamma$ (gal. center)
  & Indirect DM
   & $\Phi_{\gamma}(1) < 1.5\times 10^{-10}~\cm^{-2}~\s^{-1}$
    & {\tt GLAST} \\
 $\gamma$ (gal. center)
  & Indirect DM
   & $\Phi_{\gamma}(50) < 7\times 10^{-12}~\cm^{-2}~\s^{-1}$
    & {\tt MAGIC} \\
 $e^+$ cosmic rays
  & Indirect DM
   & $(S/B)_{\text{max}} < 0.01$
    & {\tt AMS-02} \\
\hline
\end{tabular}
\end{table}

The main results from our combined study~\cite{Feng:2001zu} of dark
matter and collider signals in supersymmetry {\em prior to the start
of the LHC} are shown in Fig.~\ref{fig:reach}.  The regions of
cosmologically interesting relic densities are shown; see
Ref.~\cite{Feng:2000gh} for details.  A generous range is $0.025\le
\Omegachi h^2 \le 1$, where the lower bound is the requirement that
neutralino dark matter explain galactic rotation curves and the upper
bound follows from the lifetime of the universe.  Above this shaded
region, $\Omega h^2 > 1$; below, $\Omega h^2 < 0.025$.  The range $0.1
\le \Omegachi h^2 \le 0.3$, preferred by current limits, is also
shown.  

Several striking conclusions follow from Fig.~\ref{fig:reach}:

\noindent {\bf No Useful Upper Bounds from Cosmology.}  For all $\tb$,
cosmologically interesting densities are possible for either
relatively small $m_0$, where the dark matter is Bino-like and
coannihilation effects may be relevant~\cite{coann}, or in the focus
point region at large $m_0 \agt 1~\tev$, where the dark matter
contains a small but phenomenologically significant Higgsino
component. As is evident from Fig.~\ref{fig:reach}, for all values of
$\tb$, the cosmologically preferred region extends to very large $m_0$
and $\mgaugino$; discovery at near future colliders cannot be
guaranteed by cosmological arguments.  The cosmologically preferred
region is ultimately cutoff around $\mgaugino \sim 6~\tev$, but this
is far beyond the reach of the LHC and planned linear colliders.

\noindent {\bf Excellent Prospects for Combined Searches.}  At the
same time, by combining all available searches, nearly all of the
cosmologically preferred models predict signals in at least one
experiment.  This is strictly true for $\tb=10$.  For $\tb=50$, some
of the preferred region escapes all probes, but this requires
$\mgaugino \agt 450~\gev$ and $m_0 \agt 1.5~\tev$, and requires
significant fine-tuning of the electroweak scale. In the most natural
regions, all models with significant neutralino dark matter will yield
some signal before the LHC begins operation.

\noindent {\bf Complementarity of Particle and Astrophysical
Searches.}  Also noteworthy is the complementarity of traditional
particle physics searches and indirect dark matter searches.  Collider
searches require, of course, light superpartners.  High precision
probes at low energy also require light superpartners, as the virtual
effects of superpartners quickly decouple as they become heavy.  Thus,
the LEP and Tevatron reaches are confined to the lower left-hand
corner, as are, to a lesser extent, the searches for deviations in $B
\to X_s \gamma$ and $a_{\mu}$.  These bounds, and all others of this
type, are easily satisfied in the focus point models with large $m_0$,
and indeed this is one of the virtues of these models.  However, in
the focus point models, {\em all} of the indirect searches are
maximally sensitive, as the dark matter contains a significant
Higgsino component.  Direct dark matter probes share features with
both traditional and indirect searches, and have sensitivity in both
regions.  It is only by combining all of these experiments, that the
preferred region may be explored completely.

\begin{figure}[t]
\includegraphics[height=2.3in]{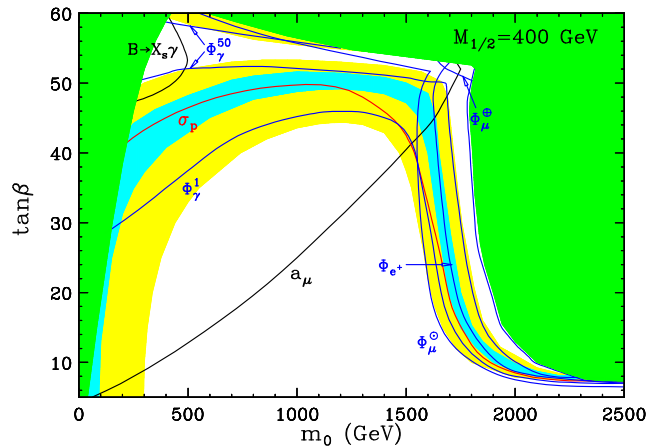}%
\caption[]{As in Fig.~\ref{fig:reach}, but in the $(m_0, \tb)$ plane
for fixed $\mgaugino=400~\gev$, $A_0 = 0$, and $\mu > 0$.  The regions
probed are toward the green regions, except for $\Phi_{\gamma}^{50}$,
where it is between the two contours.  The top excluded region is
forbidden by limits on the CP-odd Higgs scalar mass.}
\label{fig:reach400}
\end{figure}

\noindent {\bf Implications for Linear Colliders.}  Finally,
these results have implications for future colliders.  In the
cosmologically preferred regions of parameter space with $0.1 <
\Omegachi h^2 < 0.3$, all models with charginos or sleptons lighter
than 300 GeV will produce observable signals in at least one
experiment.  This is evident for $\tb=10$ and 50 in
Fig.~\ref{fig:reach}.  In Fig.~\ref{fig:reach400}, we vary $\tb$,
fixing $\mgaugino$ to 400 GeV, which roughly corresponds to 300 GeV
charginos.  We see that the preferred region is probed for any choice
of $\tb$. (For extremely low $\tb$ and $m_0$, there appears to be a
region that is not probed.  However, this is excluded by current Higgs
mass limits for $A_0 = 0$.  These limits might be evaded if $A_0$ is
also tuned to some extreme value, but in this case, top squark
searches in Run II of the Tevatron \cite{stop} will provide an
additional constraint.)  These results imply that if any superpartners
are to be within reach of a 500 GeV lepton collider, some hint of
supersymmetry must be seen before the LHC begins collecting data.
This conclusion is independent of naturalness considerations.

While our quantitative analysis is confined to minimal supergravity,
we expect this result to be valid more generally.  For moderate values
of $\tb$, if the dark matter is made up of neutralinos, they must
either be light, Bino-like, or a gaugino-Higgsino mixture.  If they
are light, charginos are typically also light, and can be discovered
at colliders.  If the dark matter is Bino-like, light sfermions are
required to mediate their annihilation, and there will be anomalies in
low energy precision measurements. And if they are a gaugino-Higgsino
mixture, at least one indirect dark matter search will see a signal.
For large $\tb$, low energy probes become much more effective and
again there is sensitivity to all superpartner spectra with light
superpartners. These conclusions are not airtight --- they may be
evaded by highly split charginos and neutralinos, highly split
sfermions, or enhanced annihilation from Higgs poles. For example, in
the case of highly split sfermions, one could arrange for light staus
to mediate neutralino annihilation, but very heavy smuons to suppress
$g_{\mu} -2$. Barring such possibilities, however, if supersymmetry
exists at the weak scale and neutralinos provide the bulk of dark
matter, there are excellent prospects for supersymmetry discovery
prior to LHC operation.


\begin{acknowledgments}
The work of J.L.F. and F.W. was supported in part by the US Department
of Energy under cooperative research agreement DF--FC02--94ER40818.
\end{acknowledgments}


\end{document}